\documentclass[11pt,preprint]{aastex}
\usepackage{natbib}
\usepackage{color}
\usepackage{graphicx}
\usepackage{rotating}
\usepackage{hyperref}
\usepackage[all]{hypcap}
\usepackage{mathrsfs}
\usepackage{amsmath}
\usepackage{amssymb}
\hypersetup{
    bookmarks=true,         
    unicode=false,          
    pdftoolbar=true,        
    pdfmenubar=true,        
    pdffitwindow=false,     
    pdfstartview={FitH},    
    pdftitle={My title},    
    pdfauthor={Author},     
    pdfsubject={Subject},   
    pdfcreator={Creator},   
    pdfproducer={Producer}, 
    pdfkeywords={keyword1} {key2} {key3}, 
    pdfnewwindow=true,      
    colorlinks=true,        
    linkcolor=blue,         
    citecolor=blue,         
    filecolor=magenta,      
    urlcolor=cyan           
}

\begin{document}
\title{Data-constrained Magnetohydrodynamic Simulation of a Long Duration Eruptive Flare}
\author{Yang Guo$^{1}$, Ze Zhong$^{1}$, M. D. Ding$^1$, P. F. Chen$^1$, Chun Xia$^{2}$, Rony Keppens$^{3}$}

\affil{$^1$ School of Astronomy and Space Science and Key Laboratory for Modern Astronomy and Astrophysics, Nanjing University, Nanjing 210023, China} \email{guoyang@nju.edu.cn}
\affil{$^2$ School of Physics and Astronomy, Yunnan University, Kunming 650050, China}
\affil{$^3$ Centre for mathematical Plasma Astrophysics, Department of Mathematics, KU Leuven, Celestijnenlaan 200B, B-3001 Leuven, Belgium}

\begin{abstract}
We perform a zero-$\beta$ magnetohydrodynamic simulation for the C7.7 class flare initiated at 01:18 UT on 2011 June 21 using the Message Passing Interface Adaptive Mesh Refinement Versatile Advection Code (MPI-AMRVAC). The initial condition for the simulation involves a flux rope which we realize through the regularized Biot-Savart laws, whose parameters are constrained by observations from the Atmospheric Imaging Assembly (AIA) on the \textit{Solar Dynamics Observatory} (\textit{SDO}) and the Extreme Ultraviolet Imager (EUVI) on the twin \textit{Solar Terrestrial Relations Observatory} (\textit{STEREO}). This data-constrained initial state is then relaxed to a force-free state by the magneto-frictional module in MPI-AMRVAC. The further time-evolving simulation results reproduce the eruption characteristics obtained by \textit{SDO}/AIA 94 \AA , 304 \AA , and \textit{STEREO}/EUVI 304 \AA \ observations fairly well. The simulated flux rope possesses similar eruption direction, height range, and velocities to the observations. Especially, the two phases of slow evolution and fast eruption are reproduced by varying the density distribution in light of the filament material draining process. Our data-constrained simulations also show other advantages, such as a large field of view (about 0.76$R_\sun$). We study the twist of the magnetic flux rope and the decay index of the overlying field, and find that in this event, both the magnetic strapping force and the magnetic tension force are sufficiently weaker than the magnetic hoop force, thus allowing the successful eruption of the flux rope. We also find that the anomalous resistivity is necessary in keeping the correct morphology of the erupting flux rope.

\end{abstract}

\textit{Unified Astronomy Thesaurus concepts:} Solar activity (1475); Solar filaments (1495); Solar prominences (1519);
Solar photosphere (1518); Solar corona (1483); Solar magnetic fields (1503)

\section{Introduction}

Magnetic flux rope eruptions in the solar atmosphere are closely related to a series of solar phenomena, such as solar flares, coronal mass ejections \citep{2011Chen}, and filament/prominence eruptions. The important physical processes include magnetic reconnection, particle acceleration, shocks, and so on. These eruptive phenomena could cause severe space weather when the eruptions are directed at the Earth. To prevent and reduce space weather disasters, we need to study the eruption mechanisms and forecast the eruptions. An effective way is to develop numerical models for solar eruptions and uncover this physical processes.

The mechanisms of magnetic flux rope eruptions are usually ascribed to magnetohydrodynamic (MHD) instabilities \citep{2000Lin,2006Kliem,2010Demoulin,2019Keppens} and magnetic reconnection \citep{1999Antiochos,2001Moore}. These two kinds of mechanisms can both be described by the MHD equations, specifically, the momentum conservation equation and the magnetic induction equation. The difference between the two mechanisms relates to the different roles of the Lorentz force in the momentum equation and the dissipation term in the magnetic induction equation in initiating the eruptions. In the scenario of MHD instabilities, it is the Lorentz force that drives the system to further deviate from equilibrium. In the scenario of magnetic reconnection, however, magnetic reconnection initially changes the magnetic field connectivity, generates new Lorentz forces, lifts magnetic structures to new equilibrium height, and then facilitates MHD instabilities. The connection between the two scenarios is that reconnection may also arise from resistive MHD instabilities, which may play a role in both linear and nonlinear MHD evolutions.

Apart from some specific processes like particle acceleration that needs a treatment of particle dynamics, the model described by the MHD equations is good enough for studying magnetic flux rope eruptions. The key is to set up an initial condition as close as possible to the real case, which is usually a nonlinear force-free field (NLFFF) model, and to provide an appropriate boundary condition. Such an MHD simulation is named data-driven \citep{2012Cheung,2016Jiang,2019Guo1,2019Hayashi,2019Pomoell} or data-constrained \citep{2013Kliem,2018Amari,2018Inoue}. In the former, the boundary condition is updated in real time, while in the latter, it is fixed to the initial value or assigned numerically. Previous studies showed that data-constrained MHD simulations could reproduce magnetic flux rope eruptions if the initial condition is constructed using observational data not much earlier than the eruption time. Therefore, in the case we lack sufficient observational data covering the whole eruption process to perform data-driven MHD simulations, we can instead perform data-constrained ones that could reproduce solar eruptions.

Coronal non-potential magnetic field is usually constructed by NLFFF models \citep{1981Sakurai,2012Wiegelmann,2014Gilchrist,2017Guo}. However, most numerical NLFFF models cannot construct magnetic flux rope structures when the bottom magnetic field is weak or the flux ropes lie high in the atmosphere. A good way to overcome this drawback is to manually insert a flux rope apriori with the help of an analytical model. The Titov--D\'emoulin model \citep{1999Titov}, which assumes that an active region comprises a major magnetic flux rope with an electric current channel and a background potential field, has been used in many pioneering efforts \citep[e.g.,][]{2004Torok,2005Torok,2017Mei}. In the Titov--D\'emoulin model, the path of the electric current channel is assumed to be a toroidal arc. Such a restriction has been relaxed in a newly developed method, the regularized Biot--Savart laws \citep[RBSL;][]{2018Titov}, which allows to prescribe an arbitrary shape of the electric current channel. \citet{2019Guo2} implemented the RBSL model in the Message Passing Interface Adaptive Mesh Refinement Versatile Advection Code \citep[MPI-AMRVAC;][]{2003Keppens,2012Keppens,2021Keppens,2014Porth,2018Xia} to study the magnetic flux rope structure of an intermediate prominence. MPI-AMRVAC is an open-source and fully documented\footnote{http://amrvac.org} numerical simulation code including multi physics models and multi numerical algorithms, powered by massive parallelization and adaptive mesh refinement capability.

There are several advantages in using RBSL to construct the initial condition for an MHD simulation. RBSL could reconstruct a flux rope in weak magnetic field regions, such as quiescent and intermediate prominence regions. RBSL could also reconstruct flux ropes high up in the corona, which are common structures usually manifested or associated with prominences, filaments, cavities, hot channels, and sigmoids \citep{2017Cheng}. Moreover, flux ropes reconstructed by RBSL and the magneto-frictional (MF) method are approximately force-free, which can be used as suitable initial conditions close to an equilibrium state before the eruption. We also note that such reconstructed flux ropes are in a non-potential state and possess large free-energy, which are close to the real cases where eruptions are prone to occur. The above advantages make the method of RBSL plus MF a good way to initiate a data-constrained MHD simulation.

Here, we make data-constrained MHD simulations for an eruption event on 2011 June 21 using initial conditions computed by the method of RBSL plus MF with varying mass density. Our primary goal is to find an optimized simulation that reproduces as many observing features as possible. Observations and modeling methods are described in Section~\ref{sec:observation}. We display an optimized model and compare it with observations in Section~\ref{sec:result}. Finally, we summarize the results and discuss the simulations in Section~\ref{sec:summary}.

\section{Observations and Modeling Methods} \label{sec:observation}
The event we model is a \textit{GOES} C7.7 class flare that occurred on 2011 June 21. The flare was initiated at 01:18 UT and peaked at 03:26 UT. An unusual characteristic of the flare is that the 1.0--8.0 \AA \ soft X-ray emission lasted until 12:00 UT as studied by \citet{2017Zhou}, suggesting that the flare is an extremely long duration event. The flare also rose slowly, which provides a good example for us to study the initial driving mechanisms of the flux rope eruption.

\begin{figure}
\begin{center}
\includegraphics[width=0.8\textwidth]{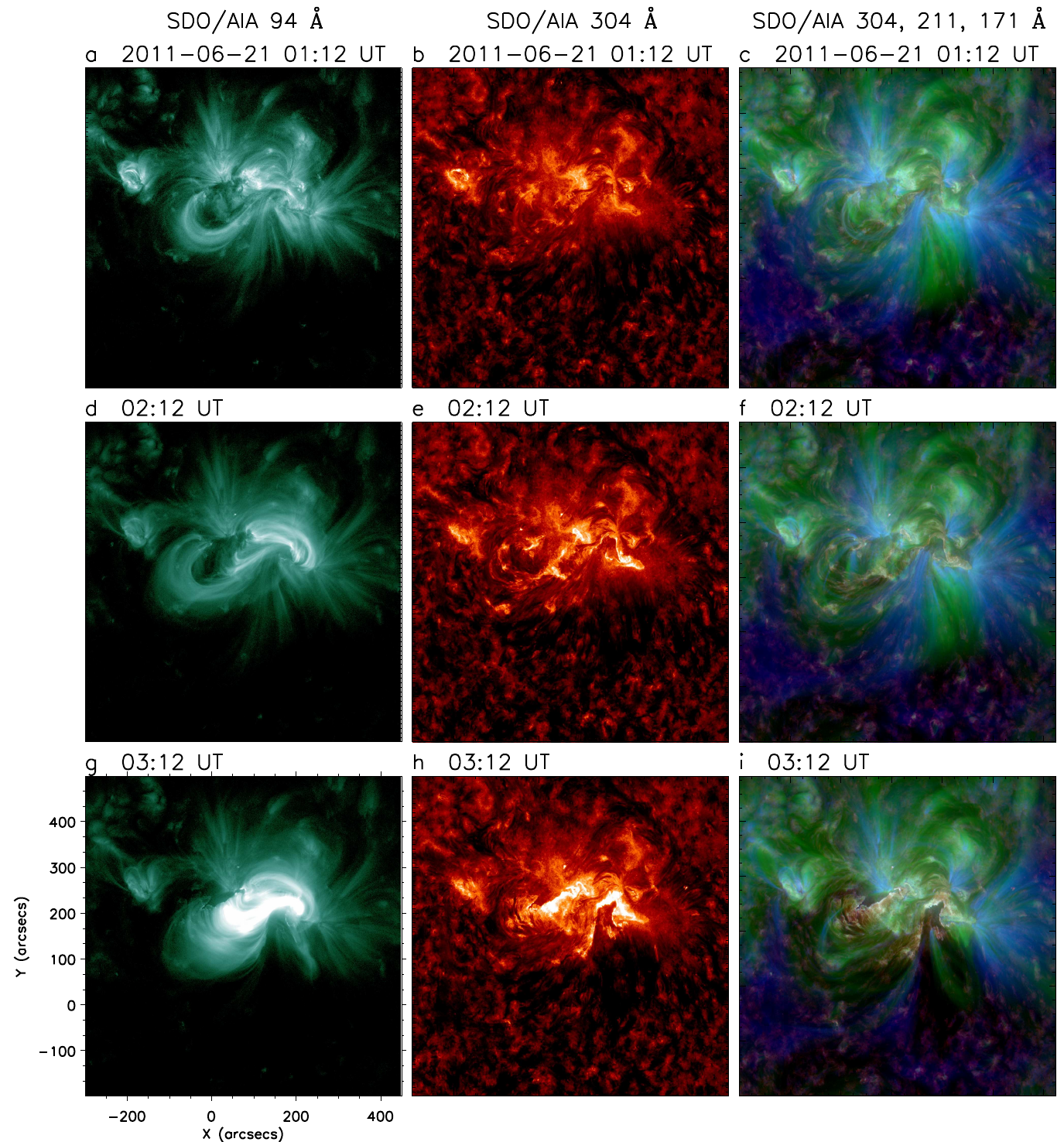}
\caption{Multi wavelength observations of \textit{SDO}/AIA showing the evolution of the flare ribbons, flare loops, and the filament. From top to bottom, the three rows display the images at 01:12, 02:12, and 03:12 UT on 2011 June 21, respectively. From left to right, the three columns display the 94 and 304 \AA \ images and composite images of 304 \AA , 131 \AA \ and 171 \AA , respectively. An animation is attached to this figure showing the time series of images in each row. The three rows in this figure show three snapshots of the animation, which cover the duration between 01:12 UT and 03:12 UT with a time cadence of 1 minute.
} \label{fig:euv}
\end{center}
\end{figure}

Figure~\ref{fig:euv} shows the evolution of the flare in multiple EUV wavelengths observed by the Atmospheric Imaging Assembly \citep[AIA;][]{2012Lemen} aboard the \textit{Solar Dynamics Observatory} \citep[\textit{SDO};][]{2012Pesnell}. The emission at 94~\AA \ of Fe \MakeUppercase{\romannumeral 18} ($\sim 6.3$ MK) reveals an expanding magnetic flux rope in the onset and rising phases at 01:12 (Figure~\ref{fig:euv}a) and 02:12 UT (Figures~\ref{fig:euv}d), respectively. The images at this wavelength also show the flaring loops  at 03:12 UT close to the peak phase (Figure~\ref{fig:euv}g). At the He \MakeUppercase{\romannumeral 2} 304 \AA \ waveband ($\sim 0.05$ MK), we can see a filament along the path of the flux rope, where the filament material flowed in the flux rope, as shown in Figures~\ref{fig:euv}b and \ref{fig:euv}e and the animation attached to Figure~\ref{fig:euv}b. The emission at this lower temperature was dominated by flare ribbons after the flare peak time. We also plot composite images by combining 304 \AA , 211 \AA \ of Fe \MakeUppercase{\romannumeral 15} ($\sim 2$ MK), and 171 \AA \ of Fe \MakeUppercase{\romannumeral 9} ($\sim 0.6$ MK) emissions that originate in a wide region spanning the upper chromosphere, the transition region, and the quiet and active-region corona, which show the evolution process of the flare (Figures~\ref{fig:euv}c, \ref{fig:euv}f, and \ref{fig:euv}i). These observations provide necessary constraints on the input parameters for the following simulations.

We select a zero-$\beta$ MHD model to simulate the evolution of the magnetic flux rope. This model neglects the gas pressure and gravity in the momentum equation, and neglects the energy equation by assuming the temperature to be zero. The basic equations are as follows:
\begin{equation}
\dfrac{\partial \rho}{\partial t} + \nabla \cdot (\rho \mathbf{v}) = 0, \label{eqn:mas}
\end{equation}
\begin{equation}
\dfrac{\partial (\rho \mathbf{v})}{\partial t} + \nabla \cdot (\rho \mathbf{v} \mathbf{v} - \mathbf{B}\mathbf{B}) + \nabla (\dfrac{\mathbf{B}^2}{2}) = 0, \label{eqn:mom}
\end{equation}
\begin{equation}
\dfrac{\partial \mathbf{B}}{\partial t} + \nabla \cdot (\mathbf{v} \mathbf{B} - \mathbf{B} \mathbf{v}) = -\nabla \times (\eta \mathbf{J}) , \label{eqn:ind}
\end{equation}
\begin{equation}
\mathbf{J} = \nabla \times \mathbf{B}, \label{eqn:cur}
\end{equation}
where $\rho$ represents the density, $\mathbf{v}$ the velocity, $\mathbf{B}$ the magnetic field, $\eta$ the resistivity, and $\mathbf{J}$ the electric current density. This set of equations is the same as that used in \citet{2020Toriumi}, but different from the one used in \citet{2019Guo1}. Here, we omit the artificial density diffusion term in Equation~(\ref{eqn:mas}) and the viscosity term in Equation~(\ref{eqn:mom}). The resistivity term in Equation~(\ref{eqn:ind}) is kept since it ensures a correct flux rope topology as shown in the following sections. Equations~(\ref{eqn:mas})--(\ref{eqn:cur}) are normalized such that the vacuum permeability, $\mu_0$, equals 1. The normalization factors for the physical variables are listed as follows: $L_0 = 1.0 \times 10^9$ cm, $t_0 = 85.9$ s, $\rho_0 = 2.3 \times 10^{-15} \ \mathrm{g \ cm^{-3}}$, $v_0 = 1.2 \times 10^7 \ \mathrm{cm \ s^{-1}}$, and $B_0 = 2.0$ G. We note that these numbers are rounded off for clarity here, while they are provided in double precision in the code and the relationships between these quantities, such as $L_0=t_0\times v_0$ and $\frac{B_0^2}{8\pi} = \frac{1}{2}\rho_0 v_0^2$, are satisfied. The resulting time step is always larger than $3.0 \times 10^{-4}$ in the normalized unit, which is a reasonable value in numerical simulations. We usually do not have the problem of negative density. But we set a lower density limit to keep the Courant--Friedrichs--Lewy (CFL) condition to a reasonable value. Otherwise, it would drop too low and make the simulation unaffordable.

The initial condition for the magnetic field is derived by combining the RBSL method \citep{2018Titov} and the MF method \citep{2016Guo2,2016Guo1}. Here, we use the same event and the same procedure as that used in \citet{2019Guo2}. There are four important parameters in the RBSL method, namely, the axis path, the minor radius, the magnetic flux, and the electric current of the flux rope. The axis path is reconstructed by triangulation from observations by \textit{SDO}/AIA and the Extreme Ultraviolet Imager \citep[EUVI;][]{2004Wuelser} aboard the twin \textit{Solar Terrestrial Relations Observatory} (\textit{STEREO}). The minor radius, $a$, of the flux rope is constrained by the width of the filament, and we adopt $a = 0.02 R_\sun = 14$~Mm in this case. The typical magnetic flux is assumed to be the average of the unsigned values on the two footprints of the flux rope within the minor radius $a$, namely, $F_0 = (|F_+| + |F_-|)/2 = 3.7 \times 10^{20}$~Mx in this event. Nevertheless, we also test different magnetic fluxes to check how they affect the eruption behavior of the flux rope in the following simulations. Finally, the typical electric current is calculated to be $I_0 = (5\sqrt{2} F_0)/(3 \mu_0 a) = 5.0 \times 10^{11}$~A. Of course, the electric current also changes if we adopt different magnetic fluxes. Figure~\ref{fig:snapshot}a shows a magnetic flux rope with $F = 3 F_0$ and $I = 3 I_0$, which represent optimized parameters to yield reasonable simulation results comparable to observations. The MHD simulation is performed in a local Cartesian coordinate system, where $x$-, $y$-, and $z$-axes point to the local west, north, and radial directions, respectively. The simulation domain is $[x_\mathrm{min}, x_\mathrm{max}] \times [y_\mathrm{min}, y_\mathrm{max}] \times [z_\mathrm{min}, z_\mathrm{max}] = [-26.5,26.5] \times [-26.5,26.5] \times [0.1,53.2]$, which is resolved by a stretched grid with  $360 \times 360 \times 180$ cells. The grid size is 0.15 per cell in the $x$ and $y$ directions, and ranges from 0.03 to 1.07 geometrically from bottom to top in the $z$ direction. We note that the field of view in the current MHD simulations is larger than that for RBSL models in \citet{2019Guo2}. The height extends to $h=53.2$ (i.e., $5.32\times 10^5$~km), which is comparable to the solar radius $6.95\times 10^5$~km. This advantage allows us to follow a flux rope to a sufficiently high altitude.

The initial condition for the density is provided by a plane-parallel atmosphere model, which is solved by $\mathrm{d} p/ \mathrm{d} z = -\rho g$ and $p=\rho T$ in the normalized form. We prescribe a temperature profile with the same form and parameters as used in \citet{2019Guo1}. The temperature profile comprises two constant values of $T_\mathrm{pho}=0.006$ below $h=0.35$ in the photosphere and $T_\mathrm{cor}=1.0$ above $h=1.0$ in the corona. They are connected by a linear function between $h=0.35$ and $h=1.0$. The normalization unit of the temperature is $T_0 = 1.0 \times 10^6$ K. It is noted that the temperature profile is only used in the derivation of the initial density profile of the atmosphere model, while it does not explicitly appear in the MHD equations of the zero-$\beta$ model (since gravity is ignored and temperature-pressure are zero by definition). The idea is to initialize the density in a realistically stratified way. The normalized density drops from $1.0 \times 10^8$ at the bottom to $0.8$ on the top of the simulation box. Since the zero-$\beta$ model allows using any density profile as the background atmosphere, we can simulate the effect of the mass inertia on the evolution of the flux rope. After a large number of tests, we find that the observed height-time profile of the filament, in particular its two phases of evolution, can only be reproduced by adopting two different density profiles. We thus split the MHD simulation into two stages, one with an initial density higher than usual in the corona and the other with an initial density given by the atmospheric model. The former case corresponds to the first evolution stage from 01:12 UT, where the initial density is $1.0 \times 10^4 \rho_0$, about 100 times the background density at the prominence height (the average value is $1.0 \times 10^2 \rho_0$). Such a case represents the added mass density within the filament. The latter case corresponds to the second evolution stage from 02:10~UT, when the filament motion starts to accelerate. We describe both cases in detail in Section~\ref{sec:kinematics}. Note that the initial velocity is zero everywhere in the computation box.

We adopt the same boundary conditions as those used in the data-constrained simulation in \citet{2019Guo1}. Namely, we assign the initial density values to two ghost layers on all the six boundaries. The velocity on these boundary layers are set to be zero during the whole simulation. The observed vector magnetic field at 01:12 UT on 2011 June 21 is assigned to the inner ghost layer of the bottom boundary. For other ghost layers, the magnetic field is determined by a zero-gradient extrapolation, where the normal component of the vector magnetic field is reset to satisfy $\nabla \cdot \mathbf{B}=0$. We note that all the physical quantities, including density, velocity, and magnetic field are assigned at the centers of the ghost cells.

Equations~(\ref{eqn:mas})--(\ref{eqn:cur}) are solved by MPI-AMRVAC. Specifically, we use a three-step method for time progressing, the HLL Riemann solver to derive the fluxes of the physical variables, and the Koren limiter to reconstruct variables from cell center to cell face. The resistivity in Equation~(\ref{eqn:ind}) is chosen to be $\eta=0$ when $J < J_{c,\mathrm{r}}$, and $\eta = \eta_0 [(J-J_{c,\mathrm{r}})/J_{c,\mathrm{r}}]^2$ when $J \ge J_{c,\mathrm{r}}$, where $J_{c,\mathrm{r}}$ is the critical current density to invoke the resistivity, and $\eta_0$ is the coefficient of the anomalous resistivity. In the following, we first present a case in Section~\ref{sec:result} with $J_{c,\mathrm{r}}=5.0$ in dimensionless units (equivalent to $8.0 \times 10^{-9}$~A~cm$^{-2}$ in cgs units) and $\eta_0 = 0.1$ (equivalent to $1.2 \times 10^{16}$~cm$^2$~s$^{-1}$ in cgs units). The simulation time ranges from 01:12 to 03:12 UT. Control parameters for this optimized Model 1 are summarized in Table~\ref{tbl1}. We also discuss some test cases by varying the control parameters in Section~\ref{sec:kinematics} and an ideal MHD case with $\eta=0$ everywhere in Section~\ref{sec:summary}.

\section{Results} \label{sec:result}

\subsection{Simulation Results and Comparison with Observations} \label{sec:comparison}

\begin{figure}
\begin{center}
\includegraphics[width=0.7\textwidth]{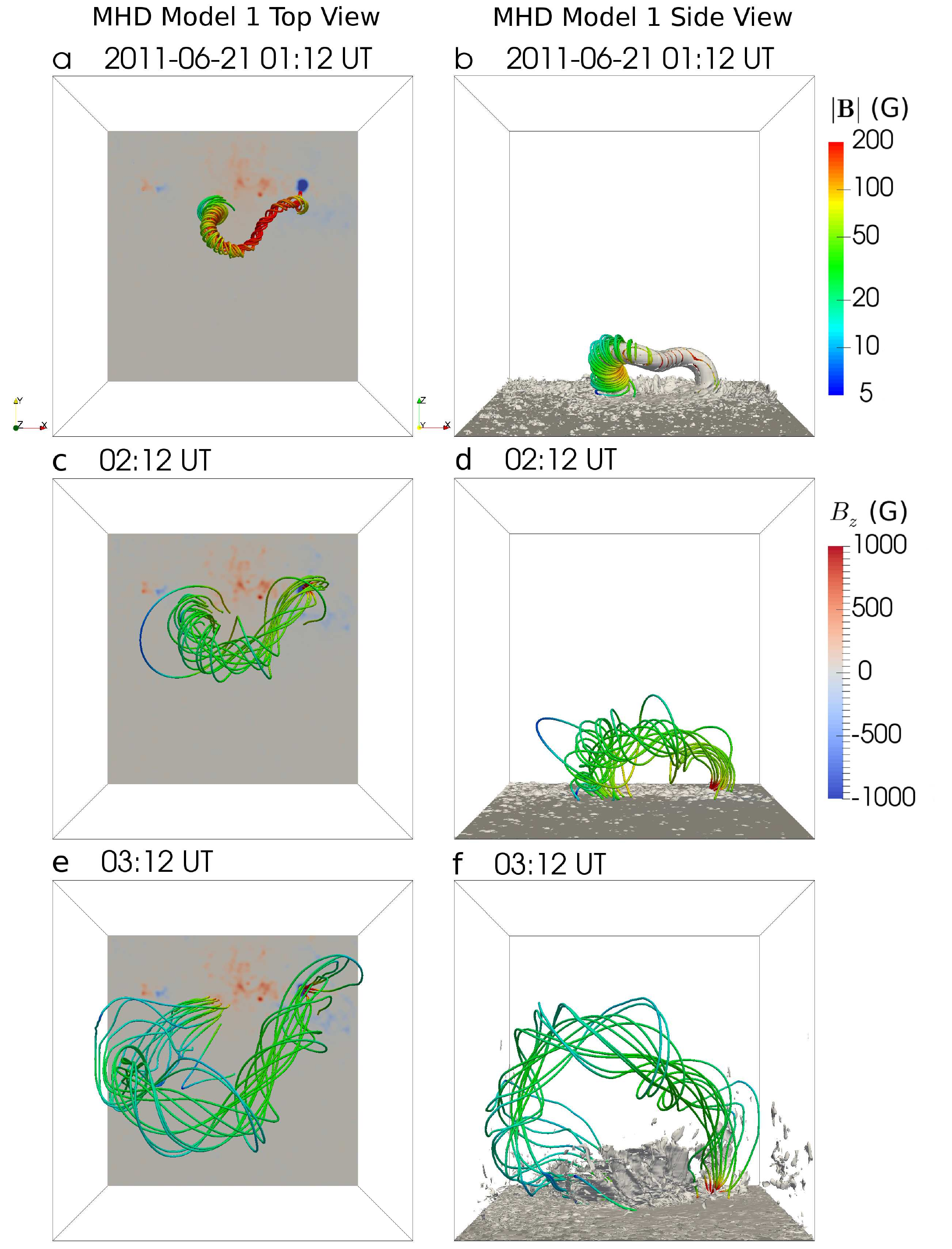}
\caption{Snapshots of an MHD simulation. From top to bottom, the three rows display the simulation snapshots at 01:12, 02:12, and 03:12 UT on 2011 June 21, respectively. The left and right columns show the evolution of the magnetic flux rope from the top and side views, respectively. The right column also shows the evolution of the electric current density, which is represented by the iso-surface at $J = 30.0$ in dimensionless units (equivalent to $4.8 \times 10^{-8}$~A~cm$^{-2}$). An animation is attached to this figure showing the time series of images in each row. The three rows in this figure show three snapshots of the animation, which cover the duration between 01:12 UT and 03:12 UT with a time cadence of 2 minutes.
} \label{fig:snapshot}
\end{center}
\end{figure}

Figure~\ref{fig:snapshot} shows the simulation results with the parameters described in Section~\ref{sec:observation}. Three snapshots in each column correspond to the flare onset, rising, and peak times, which are exactly the same times for observed images shown in Figure~\ref{fig:euv}. Animations attached to Figure~\ref{fig:snapshot} display a more detailed evolution process with a time cadence of 2 minutes during 01:12 to 03:12 UT. The flux rope evolves slowly before 02:10~UT as shown in Figures~\ref{fig:snapshot}a, \ref{fig:snapshot}b, \ref{fig:snapshot}c, \ref{fig:snapshot}d, and the attached movies. It rises and expands drastically after 02:10 UT as seen in Figures~\ref{fig:snapshot}e and \ref{fig:snapshot}f. The electric current density shows a fluctuating behavior during the whole period: it is initially high in the flux rope (Figure~\ref{fig:snapshot}b) and then dissipated to a low level in the slowly evolving process (Figure~\ref{fig:snapshot}d); however, it increases to a high level again when the flux rope starts to erupt and finally decreases again to an even lower level (watch the attached animation after 02:14~UT).

To compare the simulation results with the observations, we back-project the magnetic field vector and its location to the viewing angles of \textit{STEREO\_A/B} and \textit{SDO}, where the new $X$-, $Y$-, and $Z$-axis point to the west, north, and the observer, respectively. The back-projection is realized by three elementary rotations $\mathcal{R}_x(-B_0) \mathcal{R}_y(-L) \mathcal{R}_x(B_1)$, where the rotation matrices can be found in \citet{2017Guo}. The latitude of the disk center, $B_0$, the longitude of the active region reference point measured from the local central meridian, $L$, and the latitude of the reference point, $B_1$, for \textit{STEREO\_A/B} and \textit{SDO} are $(6.9^\circ,-90.9^\circ,14.2^\circ)$, $(-7.2^\circ,98.1^\circ,14.2^\circ)$, and $(1.7^\circ,5.6^\circ,14.2^\circ)$, respectively.

\begin{figure}
\begin{center}
\includegraphics[width=0.9\textwidth]{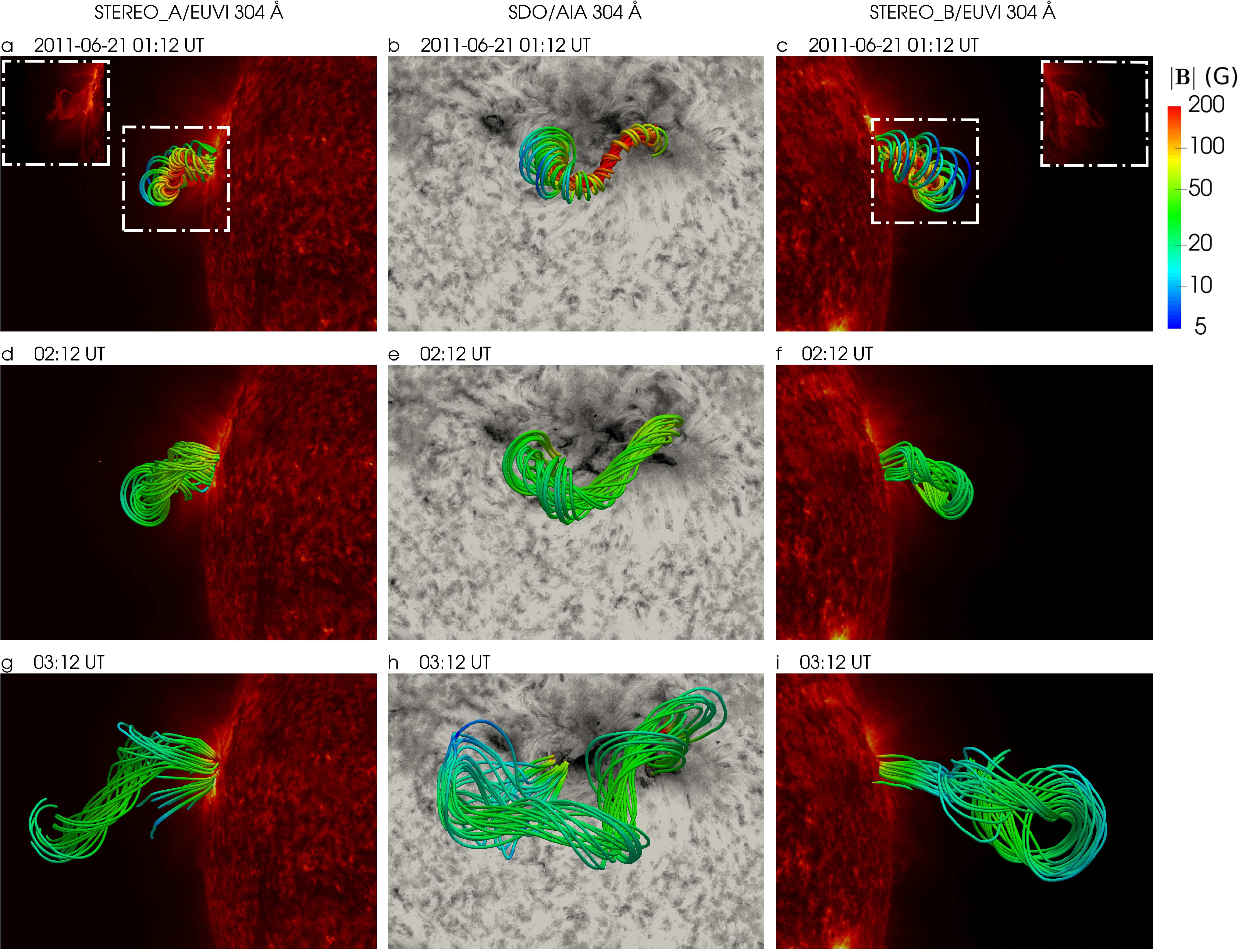}
\caption{Simulated flux ropes overlaid on the 304~\AA \ images. From top to bottom, the three rows display simulations and observations at 01:12, 02:12, and 03:12 UT on 2011 June 21, respectively. From left to right, the three columns display the images with the viewing angles from \textit{STEREO\_A}, \textit{SDO}, and \textit{STEREO\_B}, respectively. The insets in panels (a) and (c) show the background 304~\AA \ images observed by \textit{STEREO\_A} and \textit{STEREO\_B}, respectively.
} \label{fig:comparison}
\end{center}
\end{figure}

We overlay the back-projected flux ropes on the 304~\AA \ images in Figure~\ref{fig:comparison}. The morphology of the flux rope resembles that of the filament very well (Figures~\ref{fig:comparison}a, \ref{fig:comparison}b, and \ref{fig:comparison}c). The eruption path deviates from the radial direction, leaning towards the south (Figures~\ref{fig:comparison}d, \ref{fig:comparison}e, and \ref{fig:comparison}f), which is consistent with the observations obtained by EUVI aboard \textit{STEREO\_A/B}. The apex of the flux rope reaches a large altitude at 03:12 UT (Figures~\ref{fig:comparison}g and \ref{fig:comparison}i). And the flux rope keeps rising after 03:12 UT, since the velocity is still large as shown in Section~\ref{sec:kinematics}. This result is consistent with the observations of the Large Angle and Spectrometric Coronagraph \citep[LASCO;][]{1995Brueckner} C2 and C3 on the \textit{Solar and Heliospheric
Observatory} (\textit{SOHO}), which display a halo coronal mass ejection associated with the flux rope eruption. Here, we only simulate the flux rope eruption until 03:12 UT. After that, with the flux rope rising high and the filament material draining down, the flux rope becomes very weak in emission. Thus, the observations obtained by \textit{STEREO\_A/B} and \textit{SDO} provide little constraints on the simulation parameters. In the future, we expect to study a flux rope eruption that appears in the field of view of \textit{SOHO}/LASCO C2 and C3, and envisage a corresponding simulation with a much larger scale. In such a case, we can study the propagation behavior of the flux rope in more detail. Here, we only focus on the onset and initial rising phases.

We then compute the quasi-separatrix layers (QSLs) based on the magnetic field in the simulation results, and compare them with the observations. QSLs are 3D thin layers where the field linkage has a huge gradient \citep{1996Demoulin}. They delineate the boundaries of special magnetic domains, such as flux ropes, spine-fan structures, hyperbolic flux tubes, and reconnecting current sheets. QSLs are quantified by the squashing factor, $Q$, and are usually defined as regions where $Q \gg 2$ \citep{2002Titov}. We compute the value of $Q$ on two slices at each snapshot with the open-source code written by Kai E. Yang\footnote{https://github.com/Kai-E-Yang/QSL}. This code implements the method proposed by \citet{2017Scott}. Note that another similar method has been proposed by \citet{2017Tassev}.

\begin{figure}
\begin{center}
\includegraphics[width=0.9\textwidth]{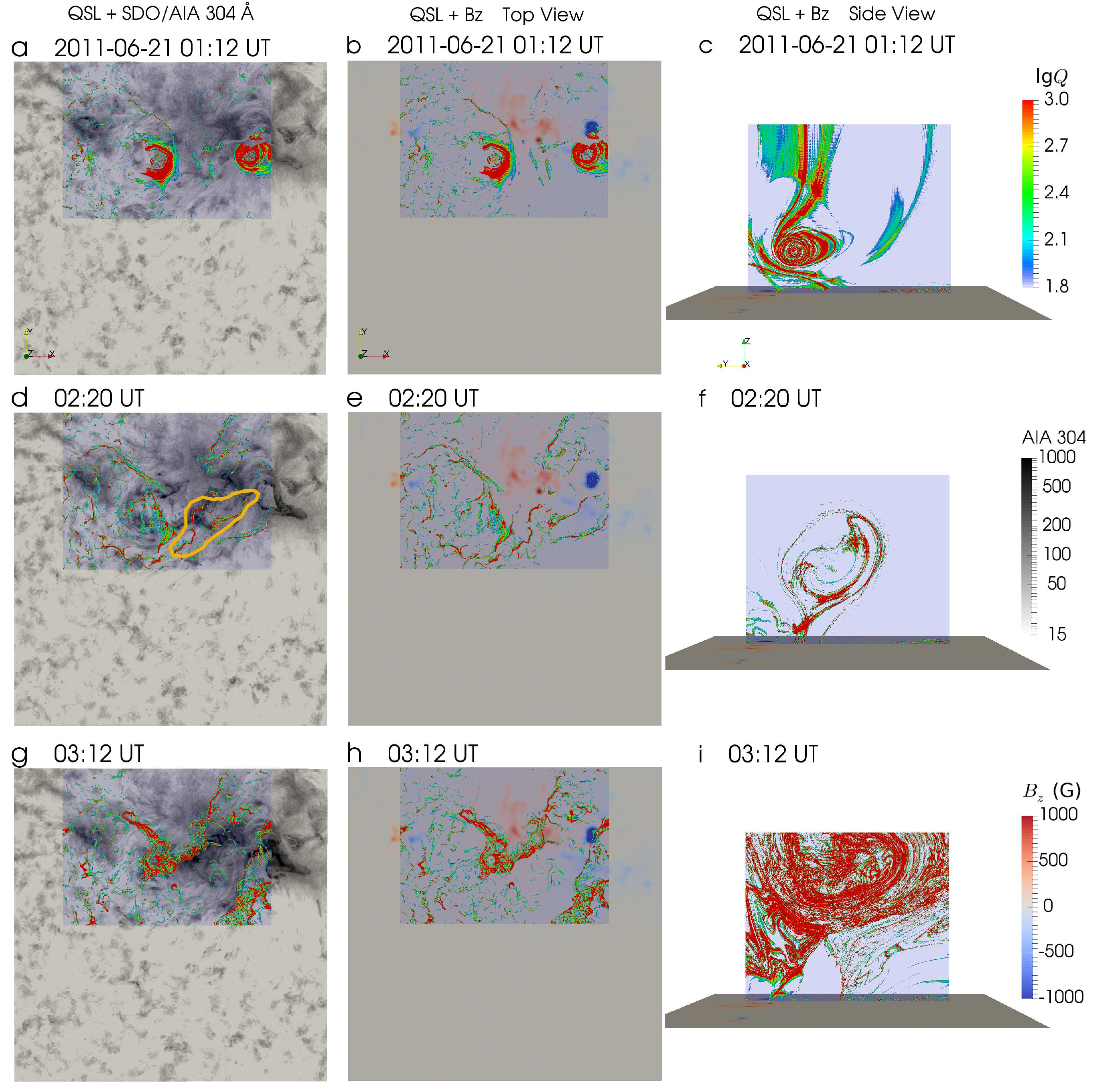}
\caption{Evolution of QSLs on horizontal and vertical slices. (a) Distributions of lg$Q$ along a horizontal slice showing QSLs (color scale) overlaid on the 304~\AA \ image (grey scale) at 01:12 UT on 2011 June 21. The slice covers the central flare region with a partial field of view of the computation domain to save computation time. (b) QSLs along a horizontal slice overlaid on the bottom magnetogram $B_z$ at 01:12 UT. (c) QSLs along a vertical slice at the central part of the flux rope at 01:12 UT. (d) Similar to panel (a) but at 02:20 UT. The orange polygon highlights two QSLs corresponding to the two flare ribbons as shown in the 304~\AA \ image. (e) Similar to panel (b) but at 02:20 UT. (f) Similar to panel (c) but at 02:20 UT. (g) Similar to panel (a) but at 03:12 UT. (h) Similar to panel (b) but at 03:12 UT. (i) Similar to panel (c) but at 03:12 UT. An animation is attached to this figure showing the time series of images in each row. The three rows in this figure show three snapshots of the animation, which cover the duration between 01:12 UT and 03:12 UT with a time cadence of 2 minutes.
} \label{fig:qsl}
\end{center}
\end{figure}

Figure~\ref{fig:qsl} displays the evolution of QSLs at three selected snapshots. The evolution of lg$Q$ distributions and 304~\AA \ images in Figures~\ref{fig:qsl}a, \ref{fig:qsl}d, and \ref{fig:qsl}g shows that the QSLs on the bottom boundary (photosphere) reproduce the flare ribbons very well, both in their shapes and evolution behavior (also see the animation attached to Figure~\ref{fig:qsl}). The most prominent feature shown in Figures~\ref{fig:qsl}a and \ref{fig:qsl}b is the QSLs surrounding the flux rope at its footprints. Figure~\ref{fig:qsl}c shows the circular QSLs surrounding and in the body of the flux rope. The time for Figures~\ref{fig:qsl}d--f is different from that for the middle panels of Figures~\ref{fig:euv}--\ref{fig:comparison}, since the QSLs coinciding with the separating flare ribbons have not fully developed at 02:12 UT. Instead, we show the QSLs at 02:20 UT in Figures~\ref{fig:qsl}d to highlight the separating QSLs within the orange polygon. A hyperbolic flux tube, where the QSLs intersect with themselves, develops under the flux rope as shown in Figure~\ref{fig:qsl}f. The QSLs coinciding with the flare ribbons further separate from each other (Figures~\ref{fig:qsl}g and \ref{fig:qsl}h), and the QSLs at the boundary and in the body of the flux rope expand drastically at 03:12 UT (Figure~\ref{fig:qsl}i). We also note that QSLs in Figures~\ref{fig:qsl}f and \ref{fig:qsl}i display some tree-ring patterns, which represent the complexity of the flux rope and the precision of the MHD simulations in this study. Both the ability to reproduce the flare ribbon evolution and the ability to resolve the complex structure of the QSLs indicate the advantage of the present MHD simulations.

\subsection{Flux Rope Kinematics} \label{sec:kinematics}

\begin{figure}
\begin{center}
\includegraphics[width=0.7\textwidth]{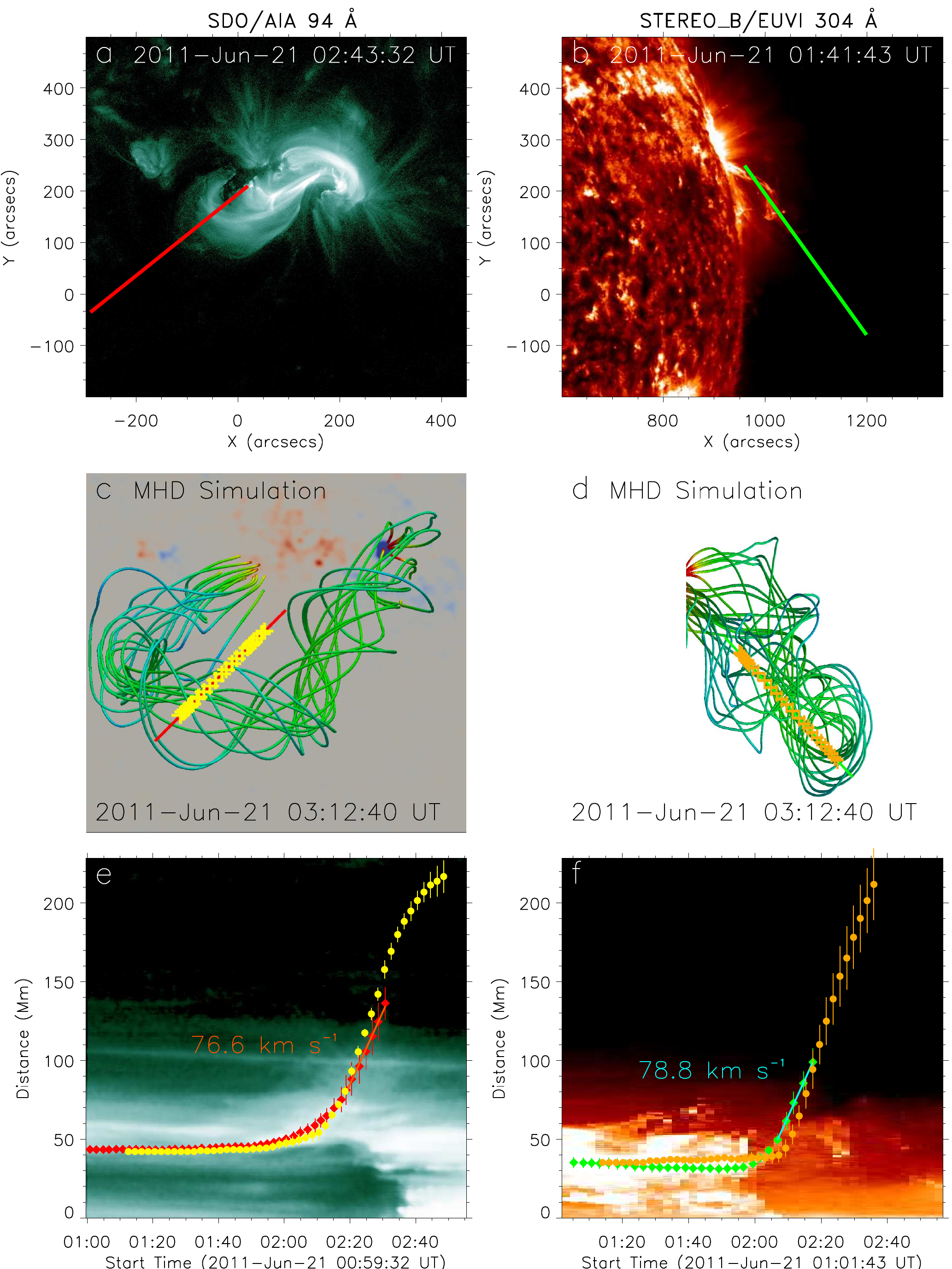}
\caption{Time-distance measurements of the filament and flux rope eruptions. (a) \textit{SDO}/AIA 94~\AA \ image at 02:43 UT. The red solid line marks the slice along which the time-distance profile is measured. (b) \textit{STEREO\_B}/EUVI 304~\AA \ image at 01:41 UT. The green solid line marks the slice for time-distance measurement. (c) A snapshot of the MHD simulation at 03:12 UT from the \textit{SDO} perspective. The red line with yellow plus signs shows an example to trace the flux rope time-distance profile along the selected slice. (d) Similar to panel (c) but from the \textit{STEREO\_B} perspective. (e) Time-distance diagram derived from the \textit{SDO}/AIA 94~\AA \ images. Red diamonds with error bars show the time-distance profile of the hot channel in observations. Yellow dots with error bars show the time-distance measurement of the flux rope in the MHD simulation. The magenta solid line is a linear fitting to the measurement at the fast eruption stage, yielding a velocity of 76.6 km s$^{-1}$. (f) Similar to panel (e), but for the measurements at the \textit{STEREO\_B} perspective. Green diamonds and orange dots are for the \textit{STEREO\_B}/EUVI 304~\AA \ observation and the MHD simulation, respectively. The cyan solid line is a linear fitting to the measurement, yielding a velocity of 78.8 km s$^{-1}$.
} \label{fig:time_slice}
\end{center}
\end{figure}

The kinematics of the flux rope obtained from the observations provides a quantitative constraint on the simulations. Therefore, we first measure the evolution of the flux rope, which is manifested by the hot channel shown at \textit{SDO}/AIA 94~\AA \ and the filament at \textit{STEREO\_B}/EUVI 304~\AA . Figures~\ref{fig:time_slice}a and \ref{fig:time_slice}b display such observations at two selected times, when the hot channel and filament are clearly observed. We select two slices along the direction of the flux rope eruption as shown in Figures~\ref{fig:time_slice}a and \ref{fig:time_slice}b. We then align the time series of the slice to get the time-distance diagrams at 94~\AA \ and 304~\AA \ as shown in Figures~\ref{fig:time_slice}e and \ref{fig:time_slice}f, respectively. From them we visually determine the apexes of the hot channel and filament. The measurements are repeated ten times to reduce the errors. We then get the average positions as shown by the red and green diamonds and the standard deviations as shown by the error bars in Figures~\ref{fig:time_slice}e and \ref{fig:time_slice}f. Most importantly, the time-distance profiles reveal two phases of evolution, namely, a slow evolution phase and a fast eruption phase. Finally, we perform linear fittings to the fast eruption phase of the time-distance profiles. The velocities are 76.6 and 78.8 km~s$^{-1}$ for the observations at 94~\AA \ and 304~\AA , respectively. Comparatively, the velocities in the slow evolution phase are small (below a few km~s$^{-1}$).

Flux rope eruptions comprising two phases of evolution are now thought to be a common phenomenon \citep{2020Cheng}. Therefore, any MHD model has to reproduce the two phases of kinematic evolution as shown in the time-distance profiles. As mentioned in Section~\ref{sec:observation}, our simulations consider two different density distributions for the two phases, a high density in the background corona (also in the flux rope) for the slow evolution phase, and a low density for the fast eruption phase. As already mentioned in Section~\ref{sec:observation}, we first determine the density distribution of the corona by solving the hydrostatic stratification. The average density in the corona at the height of the flux rope apex is about $1.0\times 10^2 \rho_0$. Since the flux rope is filled with filament material, its density is much higher than the average density in the corona. We have to artificially set a higher initial density for the simulation of the slow evolution phase. Here, we change the density to $1.0\times 10^4 \rho_0$ at the places where the density is initially lower than this value. It is worth noting that a factor 100 in density is the typical value quoted for any prominence material. At 02:10 UT, however, the density is reset to the initial value calculated from the atmosphere model.

This setup is motivated by the filament material drainage, which is universally observed before an eruption \citep{2019Jenkins,2020Chen}. Such a drainage is present in this event during the filament evolution (see the animation attached to Figure~\ref{fig:euv}). In the slow evolution phase, the magnetic flux rope lies lower in the atmosphere with higher density. When it starts to rise, filament material drains down and the mass density in the flux rope decreases. However, the zero-$\beta$ model cannot deal with the density evolution accurately. That is why we adopt two distinct initial density distributions for the two evolution phases. Figure~\ref{fig:ideal}a shows a typical snapshot for Model 1 at 02:32 UT. It is used for comparison with other test cases discussed as follows.

The eruption velocity of the flux rope is sensitively controlled by the magnetic field strength and the density within the flux rope. We have tested different cases with different magnetic fluxes of the flux rope, namely, $F=1F_0$, $2F_0$, $3F_0$, and $5F_0$, but the same initial density from the atmosphere model. The results show that the time-distance profiles are all linear in the 4 cases. The flux rope starts to rise at the beginning of the simulation, and the two phases of the eruption, as shown in the observations, cannot be reproduced. Quantitatively, the velocities in cases of $F=1F_0$ and $2F_0$ are too small to be compared to the observations in the fast eruption phase, but the velocity in the case of $F=5F_0$ is too large. Figure~\ref{fig:ideal}b shows a snapshot for the test case of $F=5F_0$ at 01:52 UT, whose control parameters are listed in Table~\ref{tbl1} as Model 2. It is found that the flux rope height of Model 2 at 01:52 UT is close to that of Model 1 at 02:32 UT, which implies that the flux rope of Model 2 erupts faster than Model 1.

We also fix the magnetic flux as $F=3F_0$ and test different density models. One group of tests is to restrict the high density region to a small area along the flux rope, roughly assumed to be the location of the filament, but reset the density to a normal value at 02:10 UT. In practice, we use the electric current density as a good physical parameter to restrict the high density region. Five cases are tested, with the critical electric current density $J_{c,\mathrm{d}} = 0.0$, 0.01, 0.1, 5.0, and 20.0. The density is initially set as $1.0\times 10^4 \rho_0$ when the current density larger than $J_{c,\mathrm{d}}$ and the background density smaller than $1.0\times 10^4 \rho_0$. When $J_{c,\mathrm{d}}=0.0$, the density in the whole corona is set to a high value; However, when $J_{c,\mathrm{d}}=20.0$, only the density in the flux rope is set to the high value. The simulation results show that the cases with $J_{c,\mathrm{d}} = 0.01$, 0.1, 5.0, and 20.0 cannot suppress the velocity to a rather small value as shown in observations before 02:10 UT. Figures~\ref{fig:ideal}c and \ref{fig:ideal}d show Model 3 with $J_{c,\mathrm{d}} = 5.0$ (see Table~\ref{tbl1} for other control parameters) at two snapshots, 01:12 UT and 01:38 UT. The density is higher than the background corona in the vicinity of the flux rope at 01:12 UT as shown in Figure~\ref{fig:ideal}c, but the flux rope has risen higher than its initial height as shown in Figure~\ref{fig:ideal}d. Note that it is supposed to be steady before the density profile is reset at 02:10 UT. With all the test parameters, we find that only the case with $J_{c,\mathrm{d}}=0.0$ could reproduce the observed time-distance profile. We also make a test by changing the time of resetting the density to normal from 02:10 UT to 01:50 UT. The results show that the fast eruption phase indeed occurs earlier correspondingly.

With the aforementioned tests, the observed time-distance profile can be reproduced in the simulation with a magnetic flux of $F=3F_0$ and a higher initial density distribution before 02:10 UT but a normal one after it. We measure the apex of the flux rope axis as shown in Figures~\ref{fig:time_slice}c and \ref{fig:time_slice}d. The measurements are repeated ten times for each view angle, one from \textit{SDO} and the other from \textit{STEREO\_B}. Then, the time-distance profiles are overlaid on that derived from observations (Figures~\ref{fig:time_slice}e and \ref{fig:time_slice}f). The errors are estimated as the standard deviation of the measurements. Therefore, by adopting appropriate parameters, the simulation can indeed reproduce both the slow evolution and fast eruption phases of the flux rope kinematics. Note that the flux rope in the simulation rises to a higher altitude than the observations could trace. The simulation thus makes up the missing story of the flux rope when it moved outside the field of view in observations.

\subsection{Eruption Mechanism} \label{sec:mechanism}

Magnetic flux rope eruptions are often thought to be driven by ideal MHD instabilities, or equivalently, the loss of equilibrium mechanism \citep{1988Demoulin,1991Forbes,2000Lin,2010Demoulin,2014Kliem}. MHD instabilities include helical kink instability \citep{2004Torok} and torus instability \citep{2006Kliem} in one single current channel, tilt-kink instability between two parallel current channels \citep{2014Keppens}, and coalescence-kink instability between two anti-parallel current channels \citep{2018Makwana}. Our simulation is most relevant to the kink and torus instabilities since there is only one major flux rope in this case. The kink instability is invoked when the twist angle is larger than a critical value, which depends on the specific equilibrium parameters, such as the magnetic field distribution, plasma $\beta$, and flux rope geometric parameters. Some typical critical twist angles for the kink instability have been proposed to be $2.5\pi$ \citep{1981Hood}, $3.3\pi$ \citep{1979Hood}, and $3.5\pi$ \citep{2004Torok}. However, it can be as large as $\sim 5\pi$ \citep{1990Mikic,1996Baty} or $6\pi$ \citep{1979Hood} for different magnetic configurations.

The torus instability occurs when the decay index, $n= - d \ln B_\mathrm{ex,pol} / d \ln {R}$, exceeds a critical value $n_\mathrm{c}$ \citep[e.g.,][]{2008Liu}, where $B_\mathrm{ex,pol}$ is the poloidal component of the external magnetic field and $R$ is the major radius of the torus-shaped flux rope. This model relies on a force balance between the outward hoop force, generated by the toroidal current and the internal poloidal magnetic field, and the inward strapping force, generated by the toroidal current and the external poloidal magnetic field. Therefore, when the external poloidal magnetic field decreases fast enough, the hoop force dominates over the strapping force, and the torus instability occurs. A classical instability analysis shows that the critical decay index, $n_\mathrm{c} = 1.5$ \citep{1978Bateman,2006Kliem}, when the current channel is semicircular and without any internal structures. It has also been suggested that $n_\mathrm{c} = 1$ if the current channel is thin and straight \citep{1978vanTend}. \citet{2010Demoulin} found that $n_\mathrm{c} \in [1.1, 1.3]$ when the current channel has a deformable shape and a finite thickness. \citet{2010Olmedo} concluded that $n_\mathrm{c}$ depends on the ratio of the flux rope apex height, $Z$, and the half distance of the flux rope footprints, $S_0$. When $Z/S_0 \ge 1$ ($Z/S_0 = 1$ is equivalent to a half torus), they found that $n_\mathrm{c} \in [0, 2]$ depending on the distribution of the external magnetic field. \citet{2010Fan} studied the conditions for the dynamic eruption of a flux rope by means of an isothermal MHD simulation. The torus instability is also regarded as the driving mechanism of the eruption and the critical decay index is found to be $n_\mathrm{c} = 1.74$ in their case.

Magnetic flux rope eruptions can be either successful or confined in the lower corona. A well-known mechanism for the confined eruption is the failed kink scenario \citep{2005Torok}, when the kink instability occurs while the torus instability does not. In this case, a flux rope undergoes a writhing motion by converting part of its twist to writhe and the eruption stops at the lower corona. \citet{2015Myers} found another mechanism to confine an eruption, namely, the failed torus scenario, which occurs when the magnetic tension force generated by the toroidal magnetic field and poloidal electric current is large enough to counterpart the outward hoop force. Large magnetic tension force requires a low twist number of a flux rope. Therefore, confinement of an eruption requires a condition that either the twist or the decay index is low, while a successful eruption is generated when both the twist and decay index are large enough. This conclusion is clearly presented in the twist (reciprocal of the safety factor) versus decay index phase diagram as proposed by \citet{2015Myers}.

\begin{figure}
\begin{center}
\includegraphics[width=1.0\textwidth]{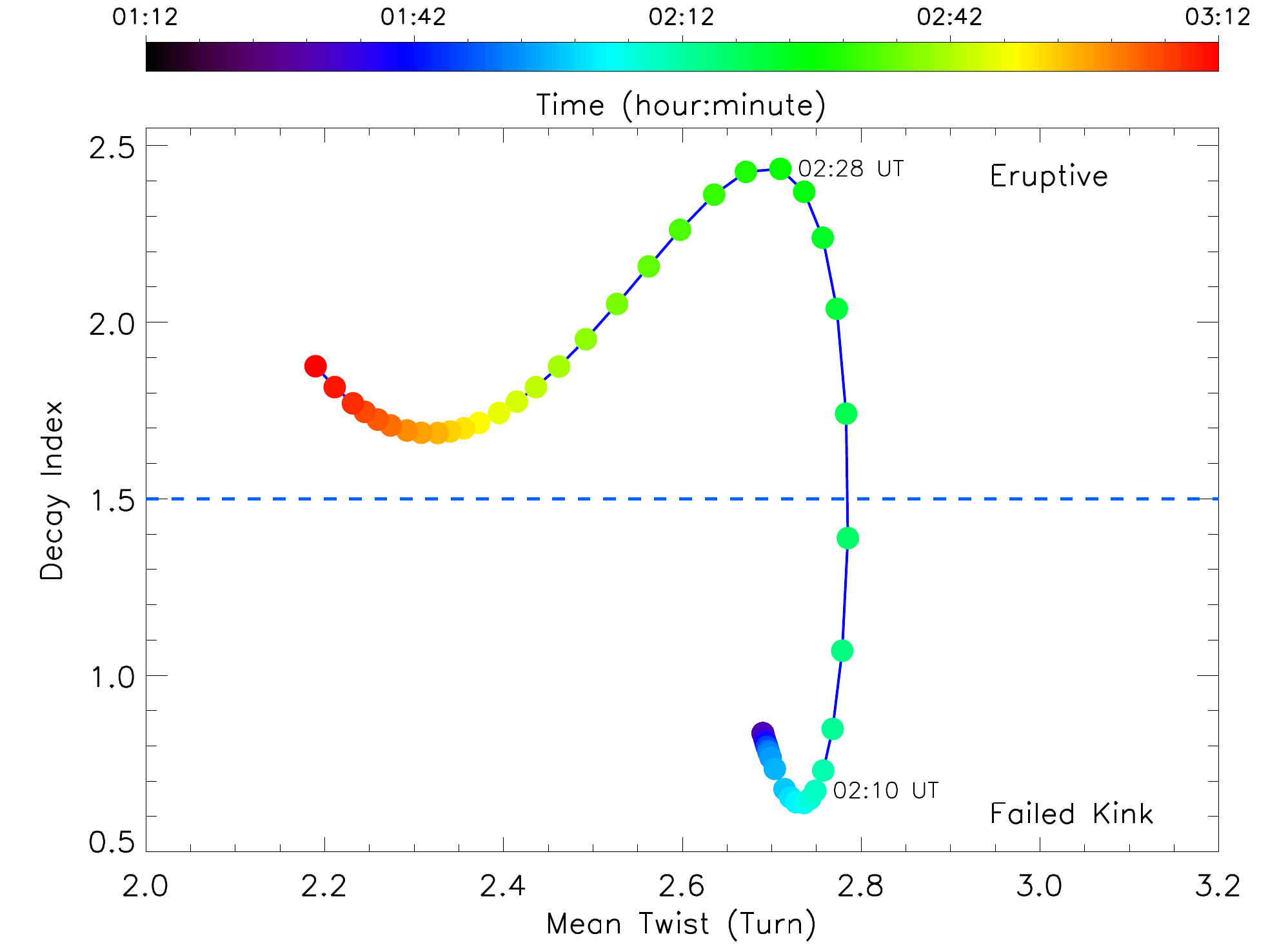}
\caption{A phase diagram showing the temporal evolution of the twist of the flux rope versus the decay index of the background magnetic field at the rope axis. Colors of the solid dots indicate the time. The dashed line marks the nominal critical value, $n_\mathrm{c}=1.5$, for the decay index. Refer to the text for more discussions on the range of $n_\mathrm{c}$.
} \label{fig:decay_twist}
\end{center}
\end{figure}

We provide the twist versus decay index phase diagram in the evolution process of the magnetic flux rope in Figure~\ref{fig:decay_twist}. The twist is computed by the Berger--Prior formula \citep{2006Berger}. Two key parameters in the calculation of the twist are the axis curve and sample field lines of the flux rope. The axis at each snapshot of the MHD models is determined as the field line most perpendicular to the vertical plane as shown in Figure~\ref{fig:qsl}b. We select 100 sample field lines randomly distributed within the border of the closed QSLs. The decay index is computed along an eruption path as shown in Figures~\ref{fig:time_slice}c and \ref{fig:time_slice}d, where the two slices determine the 3D eruption path with a unit vector of $\mathbf{e}_\mathrm{pro}$. Then, we assume the flux rope axis pointing to $\mathbf{e}_\mathrm{axi}$, which is assumed to be a unit vector perpendicular to $\mathbf{e}_\mathrm{pro}$ and parallel to the horizontal plane. There are two solutions for $\mathbf{e}_\mathrm{axi}$. We choose the one yielding positive poloidal field component. Finally, we determine the poloidal direction with a unit vector, $\mathbf{e}_\mathrm{pol}$, by requiring $\mathbf{e}_\mathrm{pol} \times (\mathbf{e}_\mathrm{pro} \times \mathbf{e}_\mathrm{axi}) = \mathbf{0}$. The poloidal component of the external magnetic field, $B_\mathrm{ex,pol}$, is determined as the scalar product of the potential magnetic field, $\mathbf{B}_\mathrm{pot}$, and the poloidal unit vector, $\mathbf{e}_\mathrm{pol}$, namely, $B_\mathrm{ex,pol}(R) = \mathbf{B}_\mathrm{pot}(R) \cdot \mathbf{e}_\mathrm{pol}$. The distance $R$ is measured along the eruption path with a start point from the bottom boundary, which is also assumed to be the major radius of the erupting flux rope. The decay index is then computed as $n= - d \ln B_\mathrm{ex,pol} / d \ln {R}$. Many previous studies only used a stationary potential field model to study the decay index \citep[e.g.,][]{2010Guo1,2012Xu,2014Zuccarello,2019Li}. Our study has both similarity and difference compared to these previous studies. We all adopt the potential field as the external magnetic field, even in the MHD simulation, since the normal magnetic field on the bottom boundary does not change too much in the eruptive process. However, in the present study, we use the evolution of the flux rope in the MHD simulation to determine the position and poloidal direction, which are important parameters to compute the decay index.

Figure~\ref{fig:decay_twist} shows that the twist varies in a small range of 2.69 to 2.75 and the decay index changes from 0.84 to 0.64 from 01:12 to 02:10 UT. Then, the twist keeps almost constant from 02:10 to 02:28 UT, while the decay index increases drastically from 0.64 to 2.37 during this period. After 02:28 UT, both the twist and decay index show a decreasing trend, except for a slight increase of the decay index in the very late phase after 02:50 UT. Figure~\ref{fig:snapshot} and the attached movies show that the flux rope does not rotate and writhe obviously during the eruption process. It indicates that the kink instability might not play a significant role in this case. Instead, the torus instability is likely the major mechanism to drive the eruption. Since the twist is high, the magnetic tension force would not be large enough to confine the eruption, which is consistent with the successful eruption of this event. Besides, as implied by our simulations in Section~\ref{sec:kinematics}, material draining due to the rising motion of the flux rope plays an important role in promoting the eruption, which reduces the inertia of the flux rope and allows it to more easily enter the torus instability regime.

\section{Summary and Discussion} \label{sec:summary}

We performed a data-constrained MHD simulation for the long duration flare of C7.7 class on 2011 June 21. The zero-$\beta$ MHD model is adopted in this study. The initial condition is provided by a magnetic flux rope structure constructed with the RBSL and MF methods. The bottom boundary is prescribed with the vector magnetic field observed at the initial time of the eruption and the velocity being fixed to zero. The simulation results reproduce the observations in several aspects. The simulated magnetic flux rope propagates along a direction deviating from the radial one and leaning towards the south, which is consistent with what is revealed by the \textit{SDO}/AIA 94 and 304 \AA \ observations. The simulated flux rope continues to rise when it moves close to the computation boundary, which indicates a successful eruption as seen from the \textit{SOHO}/LASCO observations. Particularly, we reproduce both the slow evolution and fast eruption phases of the flux rope kinematics as observed at 94 \AA \ and 304 \AA \ by assuming two different density distributions in two stages, a higher initial density in the first stage and a lower (normal) initial density in the second stage. Such a change of the initial density roughly imitates the draining process of filament material, which cannot be fully incorporated in zero-$\beta$ simulations.

The zero-$\beta$ MHD model omits the gas pressure, gravity, and the energy equation. This is acceptable if the magnetic pressure is much larger than the gas pressure and gravity, which is plausible in the solar corona. There are also some difficulties in the full thermodynamic MHD simulations, which require for very high spatial resolutions and small time steps to resolve the pressure gradient and energy evolution, making them too expensive to be affordable in simulating flux rope dynamics from the photosphere to the corona. In this sense, the zero-$\beta$ MHD model is economic and able to simulate the magnetic topology, flux rope velocity, and effects of resistivity reasonably in many circumstances. On the other hand, the thermodynamics of the plasma cannot be properly revealed in such a simplified model. Therefore, a zero-$\beta$ MHD model is unable to simulate the filament formation and dynamics in a consistent way. In our case, the mass density within the filament is supposed to be very high at the initial time in order to stabilize the magnetic flux rope for a relatively long period, an hour or so before the eruption. After the rising motion of the flux rope is triggered, the density gradually decreases due to material draining. In order to take into account such a change in the density, we adopt two different initial density values in the two stages of our simulations. Although this is a simplified treatment, our results do indicate that the material draining can play an important role in promoting the flux rope eruptions. This further confirms the finding by \citet{2006Zhou} and \citet{2020Fan}. In the future, we need a full MHD model to more accurately treat the thermodynamics of the flux rope that is omitted by the zero-$\beta$ model.

\begin{figure}
\begin{center}
\includegraphics[width=0.8\textwidth]{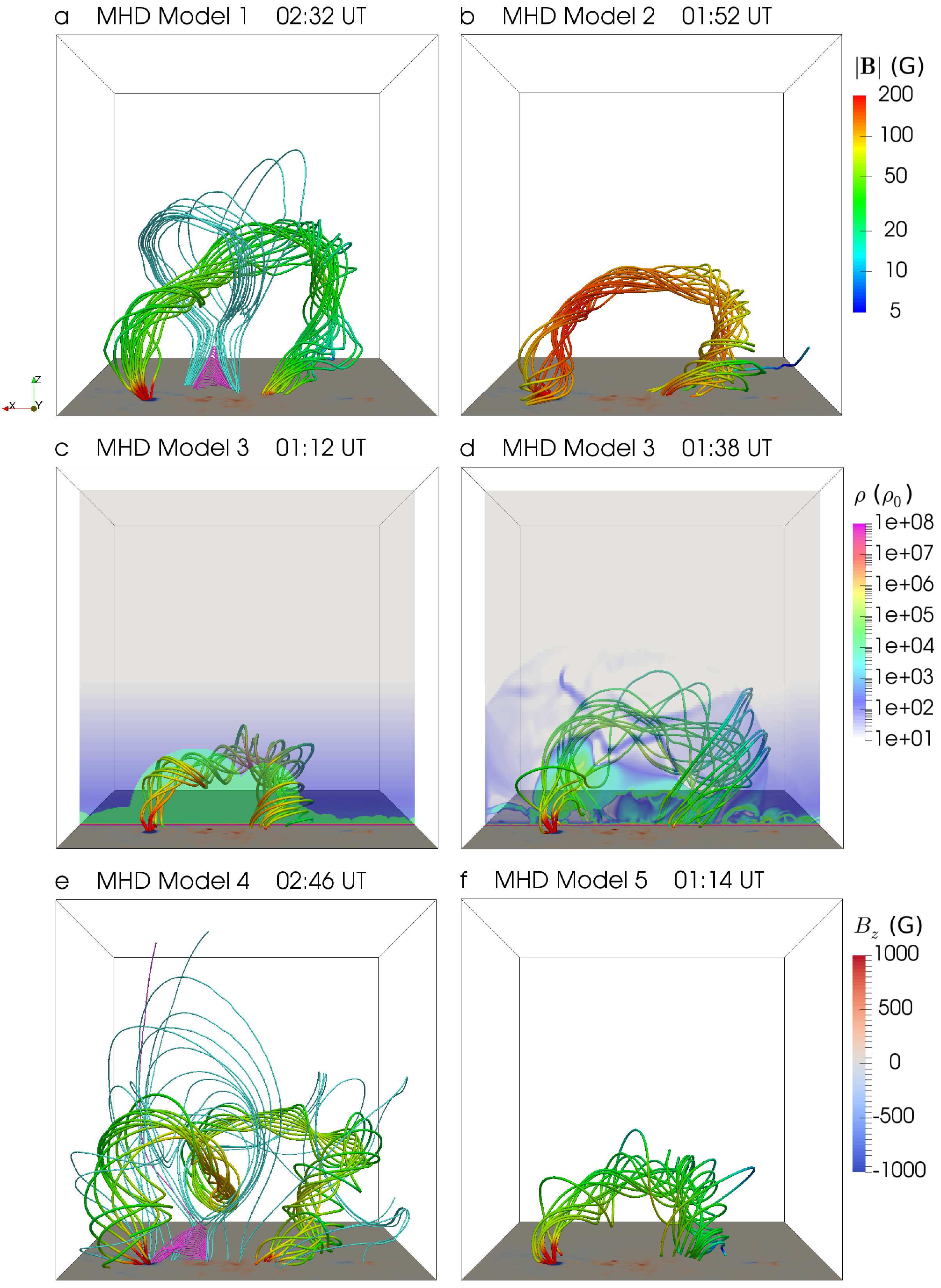}
\caption{Selected snapshots of five different MHD models viewed against the $Y$-direction. The red-yellow-green solid lines represent the field lines in the flux rope. The cyan solid lines represent the field lines stretched by the flux rope. The magenta lines represent the flare loops below the stretched current sheet. (a) A snapshot at 02:32 UT of the optimized model (Model 1). (b) A snapshot at 01:52 UT of Model 2. (c) A snapshot at 01:12 UT of Model 3. (d) A snapshot at 01:38 UT of Model 3. (e) A snapshot at 02:46 UT of the ideal MHD model (Model 4). (f) A snapshot at 01:14 UT of Model 5.
} \label{fig:ideal}
\end{center}
\end{figure}

In our simulations, the resistive term in Equation~(\ref{eqn:ind}) is determined such that $\eta=0$ when $J < J_{c,\mathrm{r}}$ and $\eta = \eta_0 [(J-J_{c,\mathrm{r}})/J_{c,\mathrm{r}}]^2$ when $J \ge J_{c,\mathrm{r}}$, where $J_{c,\mathrm{r}}=5.0$ and $\eta_0 = 0.1$. In an ideal case, the resistive term is zero everywhere, as adopted in many simulations \citep[e.g.,][]{2003Torok,2007Fan}. We also tested the ideal case by adopting $\eta=0$ everywhere. Figures~\ref{fig:ideal}a and \ref{fig:ideal}e compare the simulation results of the resistive and ideal cases (Models 1 and 4 in Table~\ref{tbl1}) at two typical times, when the eruption fully develops. It is found that the middle part of the flux rope kinks down to the bottom boundary in the ideal case (Figure~\ref{fig:ideal}e), while the flux rope erupts entirely in the resistive case (Figure~\ref{fig:ideal}a). Both cases produce current sheets under the erupting flux rope, where stretched envelope field lines of opposite directions meet and reconnect there (also refer to \citealt{1997Yokoyama}). Note that in the ideal case, unavoidable numerical resistivity controls the reconnection processes, driven by round-off and truncation errors. We have also tested a few cases by varying the parameters of $J_{c,\mathrm{r}}$ and $\eta_{0}$, say, $J_{c,\mathrm{r}}=0.1, 5.0$, and 20.0, and $\eta_0=0.001, 0.01, 0.1$, and 1.0. The results show that a large value of $J_{c,\mathrm{r}}$ ($\ge 20.0$) or a small value of $\eta_0$ ($\le 0.01$) cannot eliminate the kink in the middle part of the flux rope. On the contrary, small $J_{c,\mathrm{r}}$ ($\le 0.1$) or large $\eta_0$ ($\ge 1.0$) would dissipate the flux rope significantly. Figure~\ref{fig:ideal}f shows the result of Model 5 with $\eta_0 = 1.0$ at 01:14 UT. The evolution of the flux rope is fully controlled by the resistive time step, which is much smaller than the dynamic time step. And the flux rope is dissipated rapidly. We therefore conclude that this specific event requires an anomalous resistivity prescription as adopted, as it would otherwise show different topological evolutions. In most simulations, $\eta$ is determined empirically and changes from case to case.

To better constrain future studies, we need more observations to construct the initial magnetic field model, including the axis path, minor radius, magnetic flux, and electric current of the flux rope. We also need a complete parameter survey in order to construct an equilibrium model including the thermal conditions. Finally, full MHD simulations are required to include more physics like the gravity, gas pressure, thermal conduction, radiation loss, and background heating that are missed in the zero-$\beta$ models.


\acknowledgments

The \textit{SDO} data are provided by NASA/\textit{SDO} and the AIA and HMI science teams. The SECCHI data are provided by \textit{STEREO} and the SECCHI consortium. Y.G., Z.Z., M.D.D., and P.F.C. were supported by NSFC (11773016, 11733003, 11961131002, and 11533005) and 2020YFC2201201. C.X. was supported by NSFC (11803031). R.K. was supported by a joint FWO-NSFC grant G0E9619N and received funding from the European Research Council (ERC) under the European Union's Horizon 2020 research and innovation programme (grant agreement No. 833251 PROMINENT ERC-ADG 2018) and by Internal funds KU Leuven, project C14/19/089 TRACESpace. The numerical computation was conducted in the High Performance Computing Center (HPCC) in Nanjing University.



\clearpage

\begin{table}[h]
\caption{Magnetic Flux, Density Model, and Anomalous Resistivity of Selected Models} \label{tbl1}
\begin{tabular}{c c c c c c c} \\
\hline \hline
Model     & $F$      & Reset Density  &  $J_{c,\mathrm{d}}$     & Resistivity   & $J_{c,\mathrm{r}}$   & $\eta_0$  \\
           &         &        & \small{($1.6\times 10^{-9}$ A~cm$^{-2}$)}  &      & \small{($1.6\times 10^{-9}$ A~cm$^{-2}$)} & \small{($1.2\times 10^{17}$ cm$^2$ s$^{-1}$)} \\
\hline
1   & $3F_0$  & On  & $0.0$ & On  & 5.0 & 0.1  \\
2   & $5F_0$  & Off & NA    & On  & 5.0 & 0.1  \\
3   & $3F_0$  & On  & $5.0$ & On  & 5.0 & 0.1  \\
4   & $3F_0$  & On  & $0.0$ & Off & NA  & NA   \\
5   & $3F_0$  & On  & $0.0$ & On  & 5.0 & 1.0  \\
\hline
\end{tabular}
\end{table}

\end{document}